\documentclass[fleqn,usenatbib]{mnras}

\usepackage{newtxtext,newtxmath}

\usepackage[T1]{fontenc}

\usepackage{soul}
\usepackage{xcolor}
\setstcolor{red}

\usepackage{xcolor}
\newcommand{\res}[1]{\textcolor{black}{#1}}

\DeclareRobustCommand{\VAN}[3]{#2}
\let\VANthebibliography\thebibliography
\def\thebibliography{\DeclareRobustCommand{\VAN}[3]{##3}\VANthebibliography}

\usepackage{graphicx}	
\usepackage{amsmath}	
\usepackage[acronym]{glossaries}
\usepackage{rotating}
\usepackage{tabularray}
\newacronym{fp}{FP}{false positive}
\newacronym{tp}{TP}{true positive}
\newacronym{tn}{TN}{true negative}
\newacronym{fn}{FN}{false negative}
\newacronym{cnn}{CNN}{convolutional neural network}
\newacronym{ft}{FT}{Fourier transform}
\newacronym{qt}{QT}{Q-transform}
\newacronym{gw}{GW}{
gravitational wave}
\newacronym{tpr}{TPR}{true positive rate}
\newacronym{far}{FAR}{false alarm rate}
\newacronym{ml}{ML}{machine learning}
\newacronym{ccsne}{CCSNe}{Core-Collapse Supernovae}
\newacronym{ligo}{LIGO}{Laser Interferometer Gravitational-Wave Observatory}
\newacronym{aligo}{aLIGO}{Advanced Laser Interferometer Gravitational-Wave Observatory}
\newacronym{em}{EM}{electro-magnetic}
\newacronym{eos}{EOS}{equation of state}
\newacronym{snr}{SNR}{signal-to-noise ratio}
\newacronym{cwb}{cWB}{Coherent WaveBurst}
\newacronym{stft}{STFT}{short-time Fourier transform}

\title[GW Detection]{LIGO core-collapse supernova detection using convolution neural networks}

\author[Z. Pan et al.]{
Zhicheng Pan,$^{1}$\thanks{E-mail: steven.pan@autuni.ac.nz}
El Mehdi Zahraoui,$^{2}$\thanks{E-mail: elmehdi.zahraoui@autuni.ac.nz}
Guillermo Cabrera-Guerrero$^{3}$ \thanks{E-mail: guillermo.cabrera@pucv.cl}
and Patricio Maturana-Russel$^{2,4}$\thanks{E-mail: p.maturana.russel@aut.ac.nz}
\\
$^{1}$Electrical Engineering Department, Auckland University of Technology, Auckland, New Zealand \\
$^{2}$Mathematical Sciences Department, Auckland University of Technology, Auckland, New Zealand\\
$^{3}$School of Computer Engineering, Pontificia Universidad Cat\'olica de Valpara\'iso, Valpara\'iso, Chile\\
$^{4}$Department of Statistics, University of Auckland, Auckland, New Zealand
}

\date{Accepted XXX. Received YYY; in original form ZZZ}

\pubyear{2024}

\begin{document}
\label{firstpage}
\pagerange{\pageref{firstpage}--\pageref{lastpage}}
\maketitle

\begin{abstract}

Core-Collapse Supernovae (CCSNe) remain a critical focus in the search for gravitational waves (GWs) in modern astronomy. Their detection and subsequent analysis will enhance our understanding of the explosion mechanisms in massive stars. This paper investigates a combination of time-frequency analysis tools with convolutional neural network (CNN) to enhance the detection of GWs originating from CCSNe.  \res{The CNN was trained on simulated CCSNe signals and Advanced LIGO  (aLIGO) noise in two instances, using spectrograms computed from two time-frequency transformations: the short-time Fourier transform (STFT) and the Q-transform. The algorithm detects CCSNe signals based on their time-frequency spectrograms.} Our CNN model achieves a  \res{near 100\%}  \res{true positive rate} for CCSNe GW events with a signal-to-noise ratio (SNR) greater than 0.5 \res{in our test set}. We \res{also} found that the STFT outperforms the Q-transform for SNRs below 0.5.

\end{abstract}




\section{Introduction}
Since Sir Isaac Newton, the nature of gravity has become one of the main subjects in physics. Newton initially described gravity as a force of attraction between masses in the $17^{\text{th}}$ century. Newton's law of universal gravitation revolutionized our understanding of celestial mechanics, explaining \res{the motion of the moon and planets}. However, the Newtonian framework had limitations, especially in explaining phenomena at cosmic scales. In the $20^{\text{th}}$ century, Albert Einstein shifted our understanding of the nature of gravity by formulating the General Theory of Relativity \citep{einstein1916foundation}. This theory redefined gravity not as a force but rather as a mass curving spacetime,  \res{with} spacetime dictating the movement of the mass. General relativity predicted phenomena that Newtonian physics could not account for, such as the bending of light by gravity, and described astrophysical dynamics precisely, such as the precession of planetary orbits.  Einstein's theory also led to the prediction of \glspl*{gw} that required a century of technological revolutions to validate empirically.  \glspl*{gw} are ripples in spacetime that emerge from the universe's most violent and energetic processes, such as the mergers of black holes and neutron stars \citep{Abbot2017}. \glspl*{gw} carry information about their origins and the nature of gravity, providing a novel method for observing and understanding the universe. 

\begin{figure}
    \centering
    \includegraphics[width=0.9\linewidth]{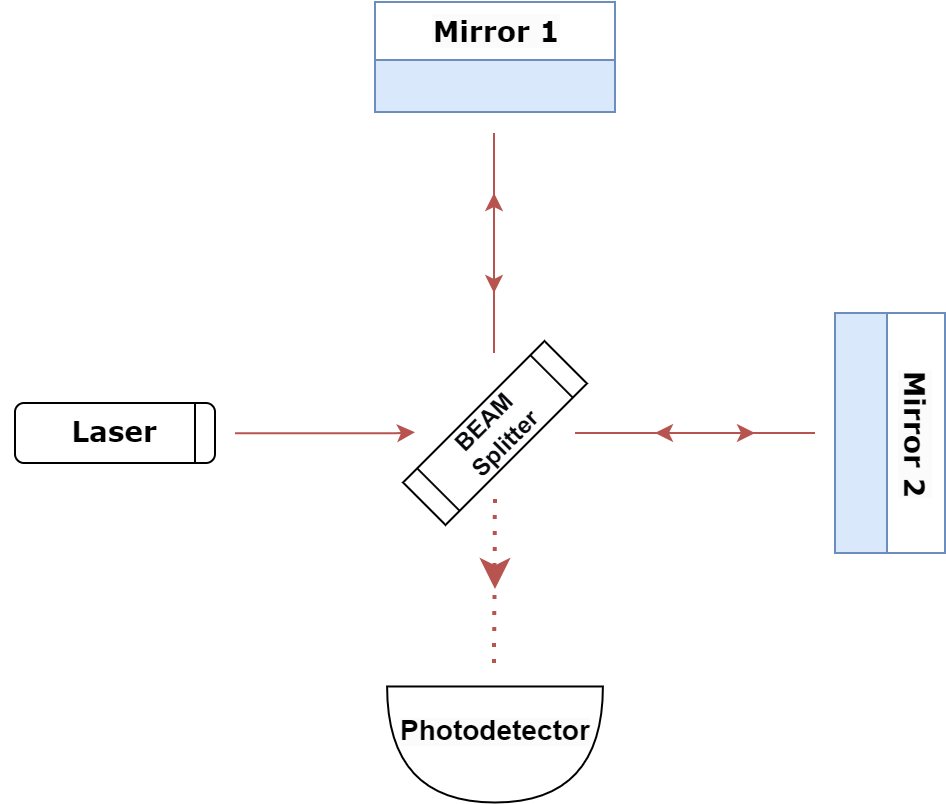}
    \caption{A simplified LIGO diagram with two perpendicular interferometry arms.} 
    \label{fig:ligo-diag}
\end{figure}

\res{The first \gls*{gw} detection, resulting from the merger of a pair of black holes with masses of approximately 36 and 29 solar masses, was announced on February 11, 2016 \citep{abbott2016ligo}.}
The event was observed on the 14$^{\text{th}}$ of September  2015 by the \gls*{aligo}'s Livingston and Hanford observatories, validating a crucial prediction of Einstein's general relativity theory and offering the inaugural direct proof of black hole mergers. LIGO detectors are GW interferometers based on Michelson's interferometer experiment. The LIGO interferometer is an optical instrument that splits a laser beam into two perpendicular arms and uses mirrors to reflect the laser beams to the photodetectors (see Figure~\ref{fig:ligo-diag}). The interference between the two laser beams will occur if the length of one of the arms or both changes, which will induce a phase shift in the laser beams. This phase shift is used to detect the changes in distance. LIGO has approximately 4 km long arms with a sensitivity of  \gls*{gw} characteristic strain up to about $10^{-23}$. The ground-based interferometer can detect \glspl*{gw} in the $200 \ \mathrm{Hz}-10 \ \mathrm{kHz}$ frequency range \citep{aasi2015advanced}. Detection of \glspl*{gw} in a lower frequency range will be possible through LISA, a space-based interferometer built to observe in the µHz-Hz frequency range \citep{danzmann2003lisa}.  LISA is scheduled to launch by 2034.  For even lower frequencies, the Pulsar Timing Array (PTA) extends the sensitivity to the nHz-µHz range, helping to understand the early dynamics of astrophysical events \citep{agazie2023nanograv}.

\gls*{aligo} has been successful over the last decade in detecting and probing the \glspl*{gw} of multiple astrophysical phenomena \citep{abbott2016gw150914,abbott2017gravitational}. However, despite being one of the main targets of these interferometers, the \gls*{gw} signals from \gls*{ccsne} have yet to be detected. A Core-Collapse supernova is a powerful and catastrophic stellar explosion that occurs at the end of a massive star's life cycle, when the core of the star, typically composed of iron, collapses under its gravity, emitting high luminosity light.  Its luminosity can be brighter than the moon's  \res{brightness} and can last for a few weeks before fading away.  This phenomenon is considered among the most powerful explosions in the universe, releasing $10^{53} \mathrm{erg}$  of gravitational binding energy. Although \gls*{ccsne} have been detected and studied across the \gls*{em} spectrum, they cannot uncover the processes deep in the star's core when the explosion ignites \citep{supernovaebook}. Therefore, studying \gls*{ccsne} through \gls*{gw} spectrum will unveil the processes contributing to this stellar explosion. In this article, we will focus on \res{detecting} \glspl*{gw} emerging from \gls*{ccsne}.

The current understanding of \gls*{ccsne} dynamics is based on two models:  the neutrino-driven mechanism and the magneto-rotational mechanism. A multi-messenger study of \gls*{ccsne} by combining \gls*{gw} and \gls*{em} \res{spectra} will help settle and evaluate our understanding of the two current models used for simulating \glspl*{gw} from \gls*{ccsne}. The two models are currently used in  two\res{-dimensional \citep{Bizouard:2021}} and three-dimensional \res{\citep{vartanyane3d}} \res{cases} to simulate the interaction between particles and generate the signatures of \glspl*{gw}.  These simulations are computationally expensive and require multiple iterations to initiate a supernova explosion \citep{muller2013new}.  \res{ A portion of   these  simulations fail to achieve an explosion state, resulting in no generation of the CCSNe \glspl*{gw} \citep{Bizouard:2021}. In \cite{wolfe2023gravitational}, a $28.3\%$ (out of $1,684$ simulations) failed to explode and no signature was obtained}. This is due to many challenges facing the \gls*{ccsne} simulation, where the resolution of the particles is the key to a successful simulation. The resolution of particles will determine how challenging the other constraints of this simulation are, such as the interactions of the electron, muon, tau neutrinos, and their anti-particles with ordinary matter \citep{Astone2018}.  The resolution used to simulate these interactions \res{also increases} the complexity of the simulation, which needs to account for the relativistic effect, sophisticated equation of state, and more constraints to achieve realistic circumstances. Generating a bank of \gls*{ccsne} waveforms is currently limited given the resources and computational power, which limits the number of \gls*{gw} templates to cover the parameter space of possible \gls*{ccsne} events \citep{wolfe2023gravitational}.  \res{The limitation of template-based matching, i.e.,} \gls*{cwb} pipeline proposed in \cite{Klimenko_2008}, has prompted the exploration of alternative methods to the template-based ones. Eventually, the new \res{alternative} methods may achieve the detection of the first \gls*{ccsne} \gls*{gw} signal in the next decade.

 Computer science has enabled the development and implementation of  \gls*{ml} algorithms, i.e., algorithms that can learn from datasets and make predictions without being explicitly programmed. These algorithms  \res{became popular} nowadays in signal and noise processing \citep{Music_CNN}.  Recently, a few \gls*{ml} techniques,  \res{particularly \glspl*{cnn}}, have been proposed and successfully tested on simulated data for the detection of \gls*{ccsne} signals \citep{Astone2018,chan2020detection}. \res{These results confirm} that employing \gls*{ml} can significantly improve the detection sensitivity of gravitational wave signals by reducing false positives and filtering out noise events.  \res{In this article, we train two \glspl*{cnn} separately using time-frequency spectrograms from the \gls*{stft} and the \gls*{qt} \citep{chatterji2004multiresolution}, computed from CCSNe GW signals and aLIGO noise data.} While \gls*{qt} has been successfully utilized for parameter inference in gravitational wave data analysis, its application for detecting  \res{CCSNe} signals represents a novel approach.  We will compare the effectiveness of these two methodologies using datasets that include various \glspl*{eos} and distances ranging from 0.1 to 10 kpc.
 

This paper is structured as follows. In Section~\ref{sec:literature_review}, we review the work done on the detection of \gls*{ccsne} \res{and discuss recent alternatives with potential in this context.} In Section~\ref{sec:methods}, we describe how the simulated data is generated and our method for processing the data and detecting \gls*{ccsne}. \res{We present and discuss the results in Section~\ref{sec:results_discussion}. Finally, we conclude by summarizing the results, discussing the limitations of the proposed methodologies, and outlining future work.}

\section{Literature Review}
\label{sec:literature_review}

Since their appearance, \gls*{ml} techniques have shown their ability to enhance and bypass many difficulties in problem-solving, especially in signal processing.  Many studies have demonstrated the possibility of training a \gls*{cnn} on time series and spectrograms to classify signals, e.g., \cite{Music_CNN}. \gls*{gw} astronomy has also incorporated machine learning techniques to enhance signal detection, especially for complex cases like searching for core-collapse supernovae signatures.   \cite{Astone2018} made the first steps in \gls*{ccsne} detection by taking advantage of the peculiarities of these \gls*{gw} signals, particularly the monotonic rise in frequency related to g-mode excitation. In their study, the simulated g-mode signature was injected into Gaussian noise to imitate the spectral behavior of LIGO. Then, the \gls*{cwb} pipeline was used to generate a time-frequency spectrogram, and a \gls*{cnn} was  \res{trained} to classify these images of spectrograms into noise and noise $+$ signal classes. This method offers a novel way to detect \glspl*{gw} from  \res{non-rotating} or slowly rotating progenitor stars, expanding the scope of detectable \gls*{gw} events.  On the other hand, \cite{chan2020detection} explored the use of \glspl*{cnn} for the classification of \gls*{gw} signals from \gls*{ccsne}. The \gls*{cnn} was tailored for multi-class classification, 
distinguishing between background noise and signals from different types of supernovae explosions magneto\res{-}rotational or neutrino-driven embedded in background noise. Additionally to \gls*{aligo} data, the \gls*{cnn} was trained with AdVirgo and KARGA, using a categorical cross-entropy loss function. The combination of four detectors allowed the authors to evaluate detections of potential extragalactic CCSNe \gls*{gw} events at $200 \ \mathrm{kpc}$. \cite{antelis2022using} marked a significant step forward in using supervised \gls*{ml} for \gls*{gw} detection from \gls*{ccsne}. The \gls*{cwb} pipeline was integrated with \gls*{ml} classifiers like linear discriminant analysis (LDA) and support vector machines (SVM). The classifiers were trained on features of the reconstructed \gls*{gw} burst, such as duration, central frequency, and detection statistics provided by the \gls*{cwb} pipeline. The classification was done independently on each of the LIGO detectors. The CCSNe models considered distances ranging  \res{from 1 to 10} kpc. Recently, \cite{nunes2024deep} tested deep learning techniques in the time domain and Neural Networks in the time-frequency domain to classify and infer on the \gls*{ccsne} parameters. The study achieved better detection levels in the time domain, reaching 98\%  for \gls*{snr} \res{greater than} $10$. These studies have demonstrated the potential of machine learning techniques to unravel the complex nature of \gls*{ccsne} and enhance the sensitivity and accuracy of \gls*{gw} detection. 

\res{The difficulty in detecting CCSNe signals lies, in part, in the need for robust signal extraction methods. Recent studies, such as \cite{Song2024a,Song2024b}, have contributed significantly to developing adaptive neural control systems in noise reduction and signal tracking, laying foundational methods that can be extended to GW analysis. Similarly, \cite{Wang2023} proposed a reinforcement learning-based strategy, namely Q-Learning, for iterative optimization in complex system identification, offering a pathway for improving the detection pipeline for faint astrophysical signals. Recent advances in neural-network-based classification techniques (e.g., \cite{Abdullahi2024a, Abdullahi2024b}) present diverse opportunities for improving CCSNe detection, such as novel strategies for effective features selection, pooling strategies to avoid over-fitting, and the use of Fourier \glspl*{cnn} for pattern detection in real-time applications, among others. The literature on algorithms with potential for CCSNe detection is extensive but remains largely unexplored.
}
\section{Methods} 
\label{sec:methods}

First, we describe the \gls*{ccsne} \gls*{aligo} \citep{aasi2015advanced} data generation process in the time domain used in this study. Then, we discuss the pre\res{-}processing of the data before generating the spectrogram images using the popular \gls*{stft}, one of the most frequently used tools in time-frequency analysis, and the  \res{\gls*{qt}}, a very popular technique in \gls*{aligo} \gls*{gw} time-frequency data analysis. Finally, we describe the \gls*{cnn} proposed for \gls*{ccsne} detection, including its architecture and the way it is trained, validated, and tested.

\subsection{Data Generation} \label{sect:DataGeneration}

\begin{figure}
    \centering
    \includegraphics[width=1\linewidth]{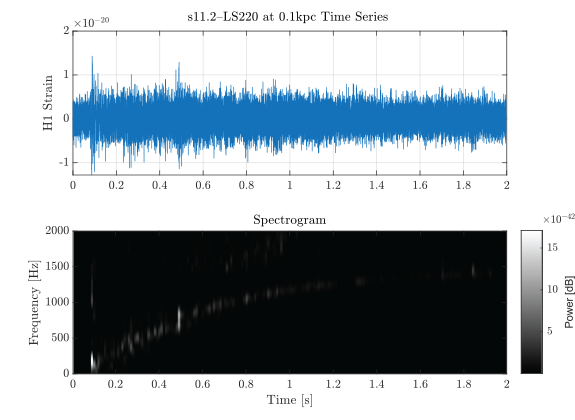}
    \caption{Top panel: CCSNe's GW signature at 0.1 kpc embedded in the \gls*{aligo} noise. Bottom panel: The \gls*{stft} spectrogram of the GW signal.} 
    \label{fig:ligo-sig}
\end{figure}

The positive class contains simulated  \res{CCSNe} \gls*{gw} signals from eight  \res{EOS models}: \textit{s11.2, s15, s20, s25, s40} under \textit{LS220}, \textit{s15} under \textit{GShen}, and \textit{s15, s20} under \textit{SFHo} where the number following the `s' denotes progenitors with zero-age main sequence mass \citep{Bizouard:2021}. The simulated waveforms are submerged by the \gls*{aligo} noise \citep{aLIGOsens:2018}, all two seconds long, pre-whitened, and sampled at $2^{14}$ Hz. We apply zero padding to the simulated waveforms that are under two seconds long. The original waveform is attenuated inversely proportional to the source distance.   \res{We consider twenty-one distances from 0.1 to 10 kpc with 0.5 kpc increment from 0.5 kpc, i.e., $\{0.1,0.5,1.0,\cdots,10.0\}$.} \res{A single realization in the time domain is presented in Figure~~\ref{fig:ligo-sig}, accompanied by its time-frequency representation obtained using the \gls*{stft}.}

\begin{figure}
    \centering
    \includegraphics[width=1\linewidth]{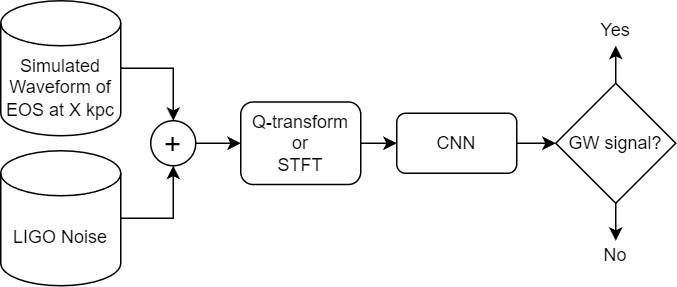}
    \caption{Data generation and analysis process. The noise data is generated by omitting the inclusion of the simulated waveform.} 
    \label{fig:data-generation}
\end{figure}

The signal is submerged in a random \gls*{aligo} noise \res{realization} generated from its power spectral density. This process is repeated 100 times for each waveform and distance. Part of the procedure is illustrated in Figure~\ref{fig:data-generation}. \res{The time-frequency transformation (either STFT or Q-transform) produces a magnitude response in the form of an image to the \gls*{cnn} input.} This procedure results in a total of  \res{100 realizations $\times$ 8 models $\times$ 21 distances $=$ 16,800 CCSNe signal} observations. These observations are divided later into training, validation, and testing datasets. The split is as follows: 5  \res{models} for training, 1 for validation, and 2 for testing. See Table~\ref{tbl:data-split} for more details.

\begin{table}
\centering
\caption{Train, validation, and test split of the data. The same number of 2,100 instances of \gls*{aligo} noise was also included for each  \res{EOS model}.}
\label{tbl:data-split}
\begin{tblr}{
  width = \linewidth,
  colspec = {Q[160]Q[94]Q[90]Q[90]Q[90]Q[90]Q[119]Q[90]Q[90]},
  cells = {c},
  column{1} = {r},
  cell{1}{2} = {c=5}{0.453\linewidth},
  cell{1}{8} = {c=2}{0.18\linewidth},
  hline{1,6} = {-}{0.08em},
  hline{2} = {-}{0.05em},
}
EOS        & LS220 &      &      &      &      & GShen & SFHo &      \\
 \res{Model}     & s11.2 & s15  & s20  & s25  & s40  & s15   & s15  & s20  \\
Train      & 2,100  & 2,100 & 2,100 & 2,100 & 2,100 & --    & --   & --   \\
Validation & --    & --   & --   & --   & --   & 2,100  & --   & --   \\
Test       & --    & --   & --   & --   & --   & --    & 2,100 & 2,100 
\end{tblr}
\end{table}

The negative class, i.e., \gls*{aligo} noise without any \gls*{gw} signals, is generated using the \gls*{aligo} power spectral density \res{\citep{aasi2015advanced}}.  The same number of 16,800 noise  \res{realizations} \res{was} generated to balance positive and negative classes.

\begin{figure}
    \centering
    \includegraphics[width=0.75\linewidth]{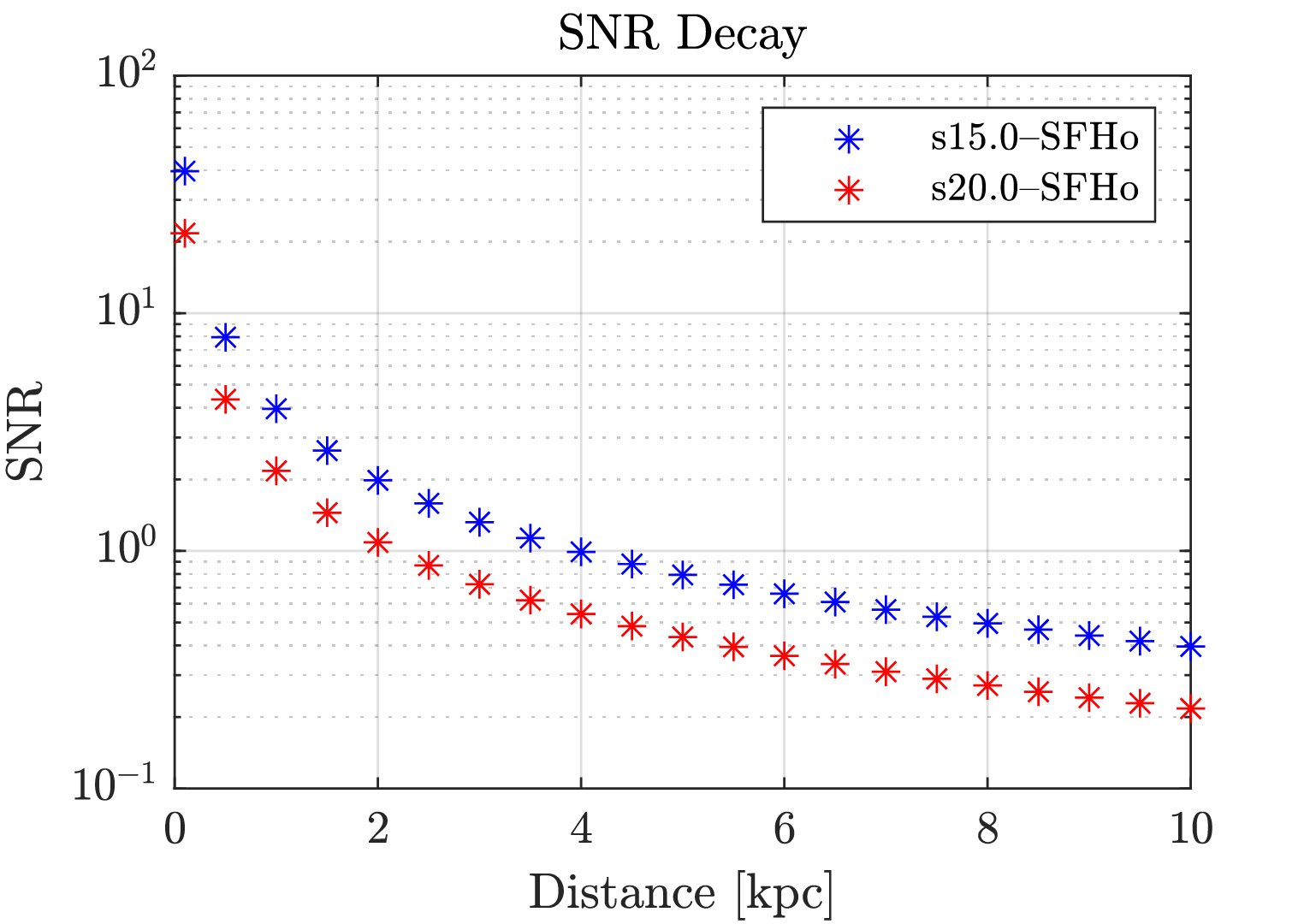}
    \caption{\gls*{snr} of two signals across distances (0.1--10 kpc).  These 2 signals are used as the test set.}
    \label{fig:snr_decay}
\end{figure}

The \gls*{snr} across distances varies for the 8  \res{EOS models}, resulting in different degrees of  \res{difficulty} for detection. To illustrate this, the \gls*{snr} of two \gls*{gw} signals against the source distance is plotted in Figure~\ref{fig:snr_decay}. The initial \gls*{snr} for \textit{s15--SFHo} and \textit{s20--SFHo} is 39.6 and 21.7 at 0.1 kpc, and the \gls*{snr} at 10 kpc is 0.396 and 0.217, respectively.  Their \gls*{snr} profiles entail a more challenging signal detection task because when compared to a similar study in \cite{Astone2018}, their smallest \gls*{snr} is 8, while our \gls*{snr} drops below 1 when the source distance is beyond 4 and 2 kpc for \textit{s15--SFHo} and \textit{s20--SFHo}, respectively.  Signals from these two  \res{EOS models} are later used as the test set in the performance measurement of the trained deep convolutional neural network for  \res{CCSNe signal} detection.

\subsection{Data Processing}\label{sect:DataProcess}


In this study, we use the \gls*{stft} and \gls*{qt} to produce the spectrograms.  The \gls*{stft} is widely used in time-frequency analyses.  We refer the reader to the extensive literature for its details and instead we focus on describing the \gls*{qt}. The \gls*{qt} has become an essential component in the \gls*{aligo} pipeline for searching gravitational wave  bursts \citep{chatterji2004multiresolution}. Initially, it was applied in the field of music signal processing to differentiate similar notes played simultaneously \citep{brown1991calculation}. The QT is a time-frequency transform designed to represent how the frequencies of a signal vary over time. This transform employs Gaussian windowed sinusoids to analyze signals, balancing time and frequency resolution. Our study uses a discrete version of the QT \citep{chatterji2004multiresolution}, incorporating a Hanning window in the frequency domain. The "Q" in Q-transform refers to the Q factor, which measures the window's width relative to its  \res{center} frequency, enabling fine-tuning of the resolution. A high Q-factor indicates a narrow window in the time domain and a broader window in the frequency domain.

The time-series signal from aLIGO, denoted as $x(t)$, is projected onto \res{$w(t-\tau,f)$} windowed complex sinusoids of frequency $f$ \res{centered around time $\tau$}, expressed mathematically \res{\citep{chatterji2004multiresolution}} as
\begin{equation}
    x(\tau,f)= \int_{-\infty}^{+\infty} x(t)w(t-\tau,f) e^{-i2\pi ft}\mathrm{d}t. 
\end{equation}


For computational efficiency, the QT can alternatively be represented using the Fourier transform of the data time-series as
\begin{equation}
    x(\tau,f)= \int_{-\infty}^{+\infty} \tilde{x}(\phi+f)\tilde{w}(\phi,f) e^{+i2\pi\phi\tau}\mathrm{d}\phi, 
\end{equation}
\res{with $\phi$ the frequency shift \citep{492555}.} This formulation allows the Fourier transform (FT) to be computed once. Subsequently, the Q-transform is calculated using this precomputed FT, applying a frequency shift and the window function in the frequency domain before inverting the FT. The QT used in our analysis is specifically adapted for aLIGO noise, normalizing the window to counteract the power spectral density of detector noise and accurately recovering the energy of transient bursts.

\begin{figure}
    \centering
    \makebox[\linewidth][c]{\includegraphics[width=1.15\linewidth]{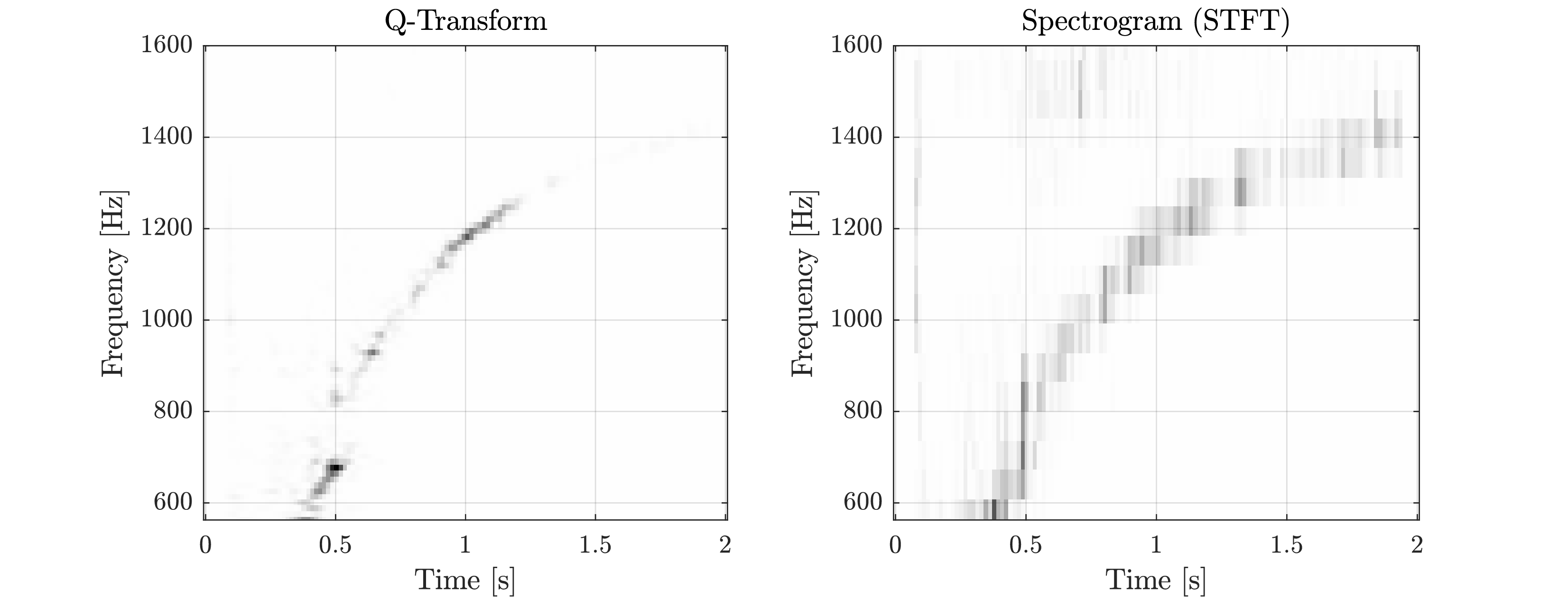}}
    
    \caption{Comparison of \gls*{qt} and \res{\gls*{stft}} spectrogram outputs of a simulated \gls*{ccsne} event, for the same frequency range. \res{The \gls*{stft} spectrogram has been truncated for comparison purposes.} Notice how the \gls*{qt} output has a higher resolution in both time and frequency \res{axes}.}
    \label{fig:q_grams}
\end{figure}

In this study, we utilize the QT implementation from the GWpy Python package \citep{gwpy} with specified parameters: \textit{qrange} 100 to 200, \textit{frange} 0 to 1600 Hz, \textit{tres} $2/128$ s, \textit{fres} $1600/128$ Hz, and \textit{whiten} = False. The term \textit{qrange} refers to the range of the Q-factor used in the QT and it is related to the time and frequency \res{spread} of the signal. A high Q-factor corresponds to a signal that is narrow in frequency but spread out in time, whereas a low Q-factor corresponds to a signal that is narrow in time but spread out in frequency. \textit{frange} is the range of frequency to compute the QT. \textit{tres} and \textit{fres} are the time and frequency resolutions of the output spectrogram from QT, respectively, and they determine the size of the spectrogram. The remaining parameters are left at their default settings. The \textit{frange} was selected based on the observation that all eight  equations of state \res{models}  studied exhibit frequency characteristics within this range. Prior to applying the QT, we band\res{-}pass the signal between 100 and 2000 Hz to attenuate irrelevant frequency components. This procedure is similarly applied to the noise input. It is important to note that this QT method interpolates the output to produce a high-resolution spectrogram on both time and frequency axes, allowing independent adjustments of \textit{tres} and \textit{fres}. Figure~\ref{fig:q_grams} demonstrates the superior time and frequency resolution of the QT compared to the spectrogram output from the short-time Fourier transform, particularly for the \textit{s11.2--LS220}  \res{model} at 0.1 kpc. The high-resolution images from the QT are directly used as inputs for the convolutional neural network, which is discussed in the following section.

\subsection{Deep Convolutional Neural Network} \label{sect:CNNMethod}

\begin{figure}
    \centering
    \includegraphics[width=0.5\linewidth]{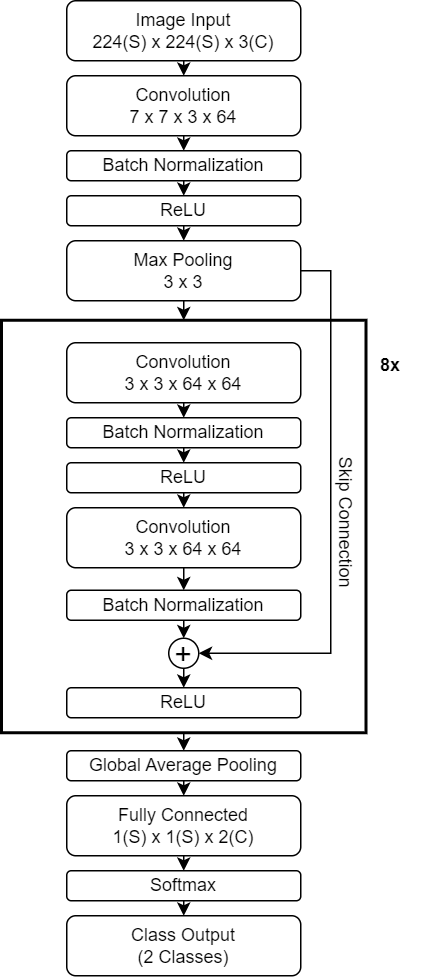}
    \caption{The architecture of the adapted deep convolutional neural network ResNet--18. The last fully connected layer is modified to suit the number of classes.}
    \label{fig:network-archi}
\end{figure}

Since our focus is on detecting the \gls*{ccsne} \gls*{gw} events from the \gls*{aligo} interferometer by using the \gls*{stft} and \gls*{qt} outputs, in 2 independent analyses, this is turned into an image classification problem. A similar work in \cite{Sassi2024} also used time-frequency spectrograms as input to classify human activity using \gls*{cnn}. We use transfer learning based on ResNet--18 \citep{He2016} as the base network, leveraging its pre-trained parameters to alleviate training challenges and harnessing its capability to capture relevant lower-level features across domains.
ResNet--18 is selected for its manageable complexity and the enhanced training efficiency afforded by its skip connections, which facilitate smoother gradient flows during backpropagation. Our network architecture, illustrated in Figure~\ref{fig:network-archi}, processes input spectrograms resized to $224 \times 224 \times 3$ (width $\times$ height $\times$ channels). The configuration follows the original ResNet--18 design, incorporating a 2D convolutional layer with a 7 $\times$ 7 kernel, batch normalization \citep{ioffe2015batch}, and ReLU activation \citep{Nair2010}. However, the final fully-connected layer is modified to have 2 neurons corresponding to the two classes of interest: event signal and the \gls*{aligo} noise. We keep the \textit{softmax} activation function to output class probabilities \res{for the positive and negative classes} $P_{s}$ and  \res{$P_n = 1 - P_s$, respectively}, that sum up to one. In the last class output layer, the prediction is made based on the positive class threshold $T$, which is between 0 and 1. If $P_s \geq T$, then the class belongs to a signal; otherwise it belongs to the noise. 

The dataset is divided into training, validation, and testing splits, as summarized in Table~\ref{tbl:data-split}. Each entry \res{in the table} has 2,100 signal instances. \res{The training and test split was performed to separate the EOSs, thus only the \textit{LS220} was used for training, \textit{GShen} for validation, and \textit{SFHo} for testing. This ensures the \gls*{cnn} trained with one EOS is able to make detections with other EOSs.} The same number of negative class instances was generated, so the trained network is not biased towards either class. To simulate the timing uncertainty inherent in actual detection scenarios and bolster the \gls*{cnn}'s robustness against overfitting, we employ random horizontal \res{image} translations up to 50\% of the image width, as part of our data augmentation \res{and neural network regularization} strategy. \res{The spectrogram $h$ is rescaled to the range between 0 and 255 as
\begin{equation}
    h_\mathrm{rescale} = 255\times\frac{h - \min(h)}{\max(h)-\min(h)}.
\end{equation}
The rescaled spectrogram is then treated as a single channel gray scale image input to the \gls*{cnn}.}

\begin{figure}
    \centering
    \makebox[\linewidth][c]{\includegraphics[width=1.15\linewidth]{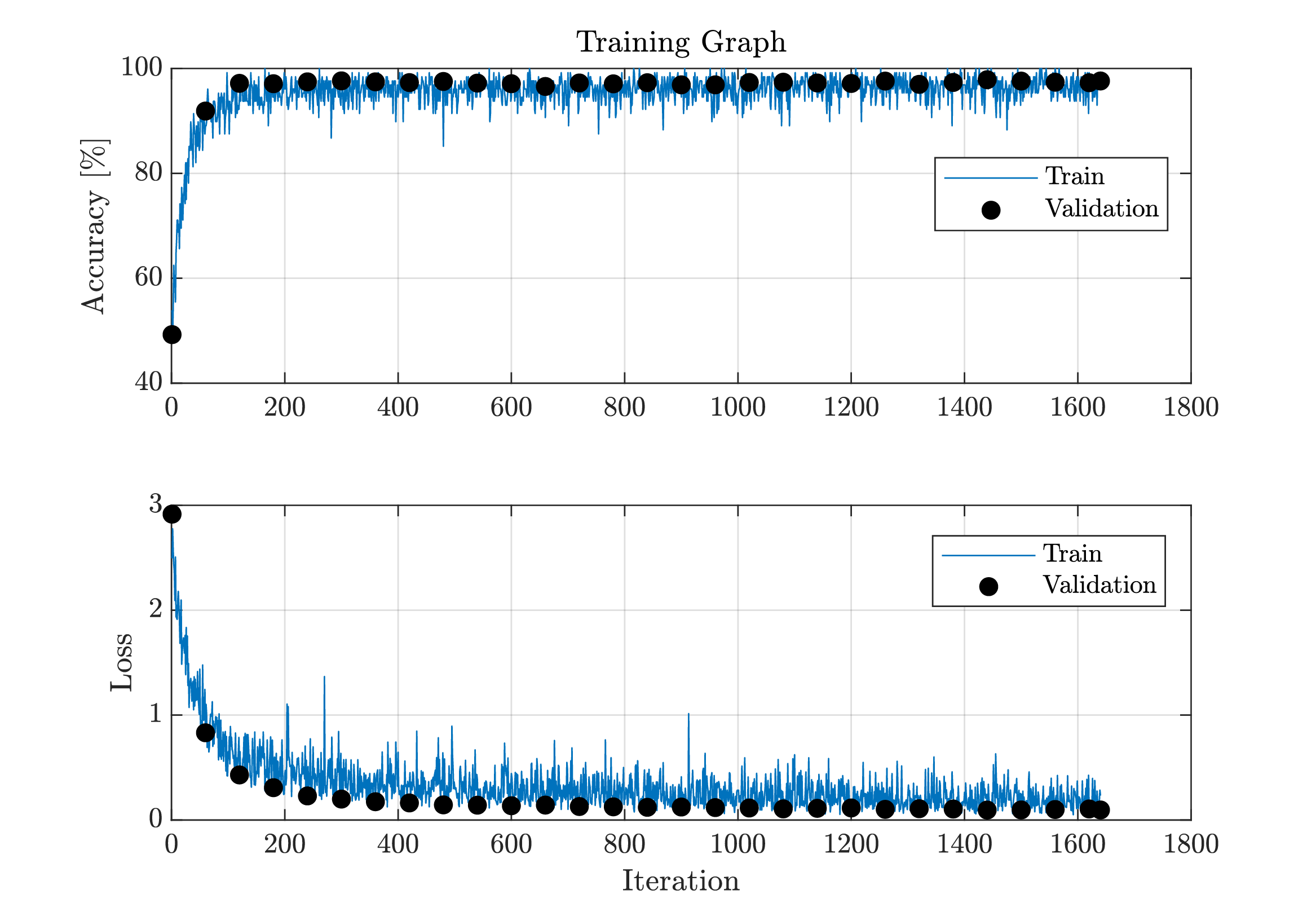}}
    \caption{Accuracy and loss plot of the training process \res{of QT-CNN}, with validation data \textit{s15--GShen}.}
    \label{fig:training_plot}
\end{figure}

For training, we employ the Adam optimizer \citep{kingma2017adam} with a learning rate of $10^{-6}$, applied over 10 epochs, and include $L_2$ regularization set at 0.05. The network, which trains \res{mini-}batches of 128 \res{training samples}, does not freeze any layers, allowing all \res{layers} to update during training. \res{These hyper-parameters were tuned such that small and similar training and validation losses were obtained.} This setup achieved a final validation accuracy of 97.95\%, \res{which is very close to the final training accuracy.} \res{This} indicates a successful optimization and learning generalization. Training progress, represented in Figure~\ref{fig:training_plot}, plots classification accuracy and network loss over training iterations. The training converges in 400 iterations. We utilize weighted binary cross-entropy loss to penalize false negatives more  \res{heavily}, enhancing the network's sensitivity to genuine GW signals. This is given by
\begin{equation}\label{eq:loss}
\mathrm{Loss} = - \frac{1}{N} \sum_{i=1}^{N} \left( w_p \cdot y_i \cdot \log(P_{s,i}) + w_n \cdot (1 - y_i) \cdot \log(1 - P_{s,i}) \right),
\end{equation}
where $w_p=5$ and $w_n=1$ represent the weights for positive and negative class errors, respectively, with $N$ the \res{mini-}batch size at 128, \res{$y_i$ the binary indicator of whether the signal instance is positive (1 for CCSNe signal) or not (0 for aLIGO noise)}.  \res{The ratio of weights $w_p$ to $w_n$ indicates that we are penalizing false negative classifications more, where a true CCSNe signal is misclassified as aLIGO noise.}

For the \gls*{cnn}  \res{trained using the spectrograms produced by QT, which we call QT-CNN,} the positive class threshold $T$ is computed by maximizing the difference between \gls*{tpr} and \gls*{far}, which means maximizing the \gls*{tpr} and minimizing the \gls*{far}. The relationship is plotted in the left of Figure~\ref{fig:TPR-FAR}, and it \res{indicates that} when $P_s \geq 0.4$\res{,} the input is classified as a  \res{CCSNe signal}. Therefore, for  \res{the QT-\gls*{cnn}, the threshold is} $T=0.4$. The \res{classification} threshold for the \res{CNN}  \res{trained using spectrograms produced by STFT, which we call STFT-CNN,} is \res{set} by default \res{at} $T=0.5$.

After hyper-parameters \res{tuning}, the network was trained again with training and validation data combined. The classification results \res{from the test set} are presented and discussed in the following section.

\section{Results and Discussion}
\label{sec:results_discussion}

\begin{figure}
    \centering
    \includegraphics[width=1\linewidth]{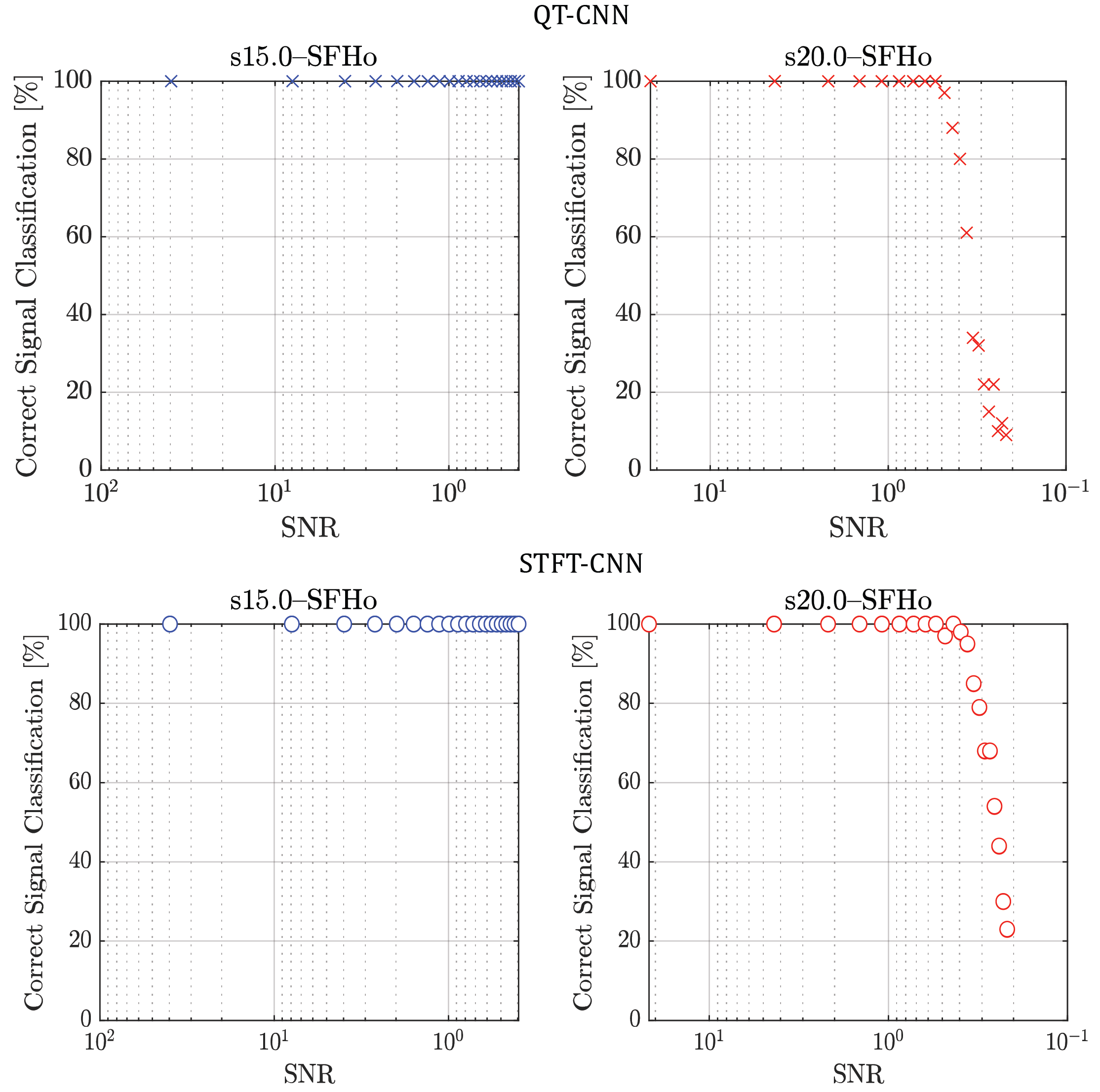}
    \caption{\res{Top:} True positive rate for \textit{s15.0--SFHo} and \textit{s20.0--SFHo} against \gls*{snr} by using the \gls*{qt} as the input to the \gls*{cnn}. There are 2100 test images for each EOS  \res{model}, positive class threshold $T=0.4$. \res{Bottom:} True positive rate by using the \gls*{stft} spectrogram as the input, positive class threshold $T=0.5$.}
    \label{fig:test_set_accuracy}
\end{figure}

The  \res{output spectrograms} from \gls*{qt} and \gls*{stft} have been used as the \res{inputs} to train the \glspl*{cnn}. \res{We denote the \glspl*{cnn} as QT-CNN and STFT-CNN, respectively.} The training procedures are the same for both models. \gls*{tpr} is first tested to get a quick overview of the model's ability to classify the signal correctly. It is plotted in Figure~\ref{fig:test_set_accuracy}  \res{for the two signals in the test set}: \textit{s15.0--SFHo} and \textit{s20.0--SFHo}.  It is expressed as the percentage of correct signal classifications over the total number of signal instances at a particular distance. In the QT\res{-\gls*{cnn}} case, for \textit{s15.0--SFHo}, all signals are correctly predicted up to 10 kpc\res{,} where the \gls*{snr} is just below 0.5. A steady, high true positive rate is observed for \textit{s20.0--SFHo} at \res{an} \gls*{snr} \res{greater than} 0.5\res{,} where the distance is 4 kpc. Beyond this \gls*{snr}, the performance starts to drop as expected, which is not surprising considering the \gls*{snr} at 10 kpc is just above 0.2. It can be observed that regardless of the \gls*{eos}, the network is capable of correctly identifying the event signal at a high success rate for an \gls*{snr} as low as 0.5. When comparing our result to that of \cite{Astone2018}, we achieved a \gls*{tpr} of 100\% for an \gls*{snr} as low as 0.5 from the \textit{s15.0--SFHo}  \res{model} where they also achieved the same \gls*{tpr} but at an \gls*{snr} of 20 \res{(interpolated from Fig. 3a)}. In the STFT\res{-\gls*{cnn}} case,  \res{the \gls*{cnn}} performs equally well for the \textit{s15.0--SFHo}  \res{model} but outperforms \gls*{qt}\res{-\gls*{cnn}} slightly for the \textit{s20.0--SFHo}  \res{model} at low  \res{\glspl*{snr}}, and achieves a \gls*{tpr} just above 20\% for \res{an} \gls*{snr} of approximately 0.2.

\begin{figure}
    \centering
    \includegraphics[width=1\linewidth]{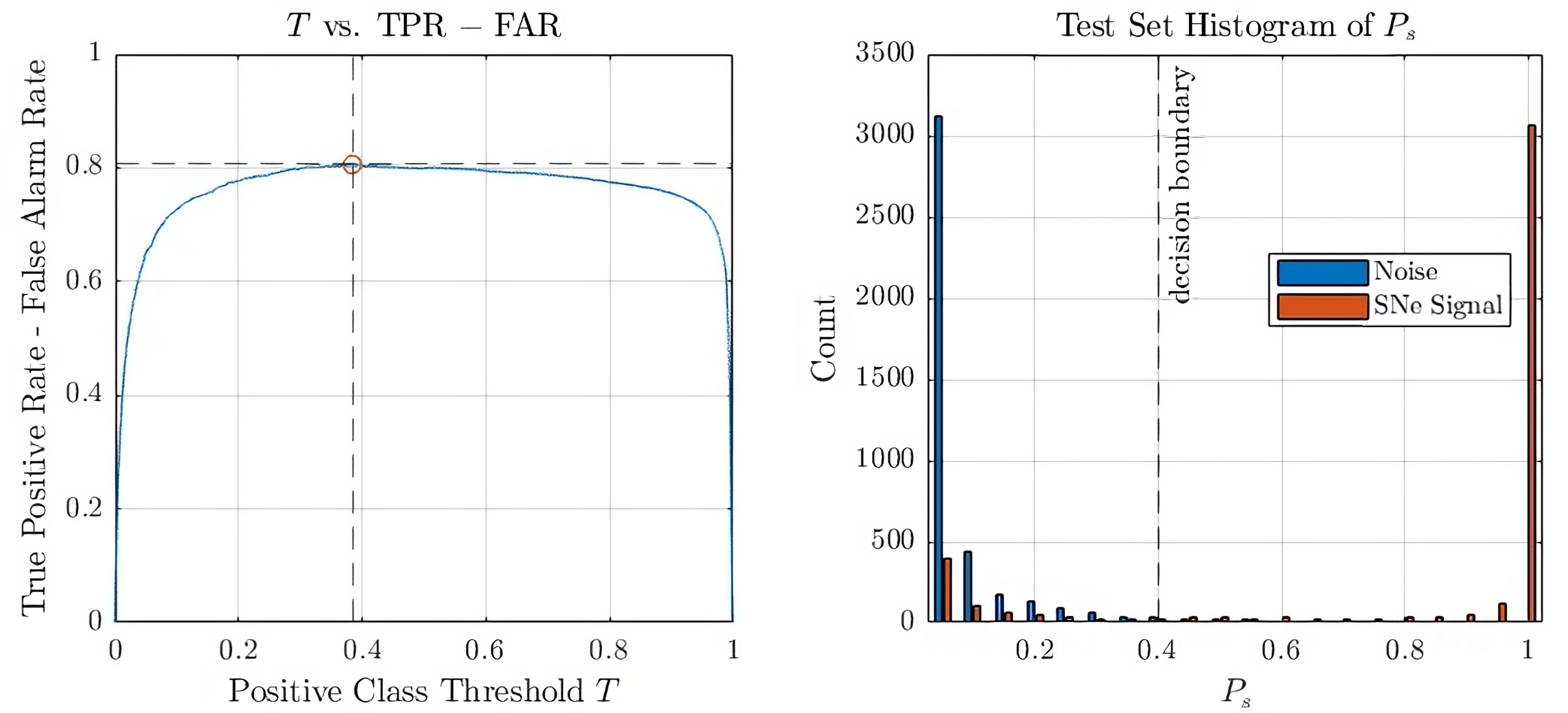}
    \caption{\res{Left:} Line plot of positive class threshold $T$ against TPR $-$ FAR for the \gls*{qt}\res{-\gls*{cnn}} case. The red marker at $T \approx 0.4$ denotes value of $T$ that maximizes TPR while minimizes FAR. \res{Right:} Histogram of $P_s$ for all test set images.}
    \label{fig:TPR-FAR}
\end{figure}

\begin{figure}
    \centering
    \includegraphics[width=0.8\linewidth]{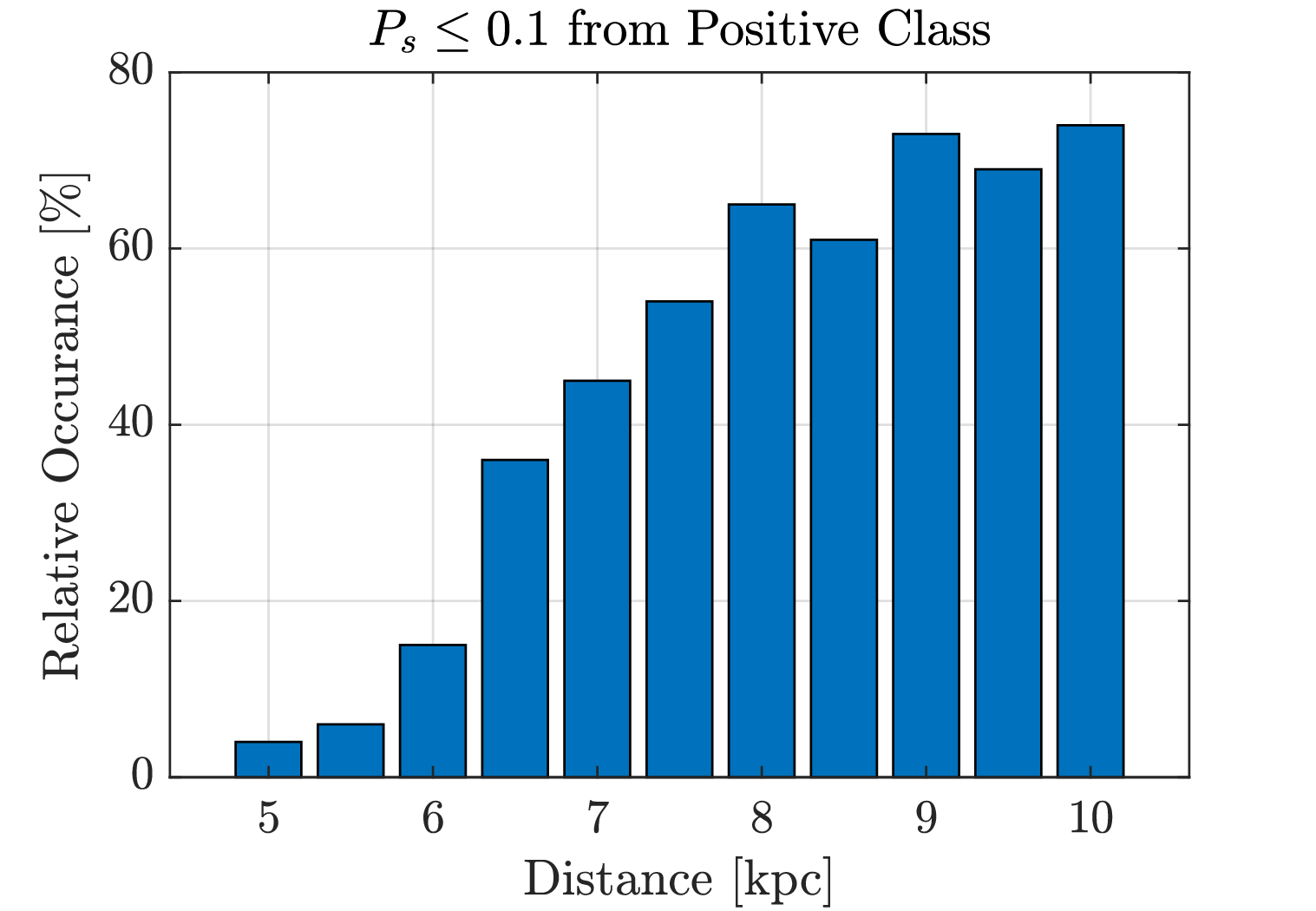}
    \caption{Relative  \res{occurrence} of test samples from \gls*{qt}\res{-\gls*{cnn}} whose $P_s \leq 0.1$ from the \textit{s20.0--SFHo}  \res{model}.}
    \label{fig:small_prob}
\end{figure}

To further analyze the network's confidence in classifying these two classes, a histogram of all positive class probabilities $P_s$ is shown on the right of Figure~\ref{fig:TPR-FAR}. The histogram indicates a clear separation of the classes, with most instances of each class distributed at opposite ends of the histogram. This observation  \res{is in agreement with} the insensitivity of \res{the} change  \res{in} $T$ to \gls*{tpr} $-$ \gls*{far} around the middle plateau, suggesting a stable binary classification model. However, it is worth noting that there are a number of  \res{false negative} samples at $P_s \le 0.1$ consisting mostly of  \res{signals} from beyond 6 kpc in the \textit{s20.0--SFHo}  \res{model}.  \res{A bar graph of their relative occurrence is plotted in Figure~\ref{fig:small_prob}, confirming the expected behavior: as the distance increases, the \gls*{snr} decreases inversely proportionally, causing the detection rate to drop.}

The overall test set performance for all source distances is summarized in the confusion chart presented in Table~\ref{tab:confusion-chart}.  The values are expressed as percentages of the ground truth. \res{We used the following metrics to evaluate the performance of the CCSNe GW signal classifier:
\begin{subequations}
\begin{align}
&\mathrm{True\ Positive\ Rate\ (TPR)} = \frac{\mathrm{TP}}{\mathrm{TP}+\mathrm{FN}}, \label{eq:TPR} \\
&\mathrm{False\ Alarm\ Rate\ (FAR)} = \frac{\mathrm{FP}}{\mathrm{FP}+\mathrm{TN}}, \label{eq:FAR} \\
&\mathrm{Precision} = \frac{\mathrm{TP}}{\mathrm{TP}+\mathrm{FP}}, \label{eq:precision} \\
&F_1\mathrm{\ score} = 2 \times \frac{\mathrm{precision} \times \mathrm{TPR}}{\mathrm{precision} + \mathrm{TPR}}. \label{eq:F1}
\end{align}
\end{subequations}
} \res{For more information about these metrics, see, for example, \cite{FAWCETT2006861}.} A \gls*{tpr} of 82.9\% and 91.5\% were achieved for  \res{QT-CNN} and  \res{STFT-CNN,} respectively, over \res{an} \gls*{snr} range from 39.6 to 0.2. For the  \res{QT-CNN}, out of 4,200 \gls*{aligo} noise and true event signal instances, 4,092 were correctly rejected as noise, and 3,482 we correctly identified as true event signals. This is equivalent to an overall true negative rate of 97.4\%, a \res{\gls*{tpr}} of 82.90\%, a \gls*{far} of 2.6\%, \res{a precision of 0.970, and a $F_1$-score of 0.893}. The \gls*{tpr} of 82.90\% is higher than the 69\% obtained in \cite{nunes2024deep}, despite our much lower \gls*{snr} of 0.1 at 10 kpc, compared to their \gls*{snr} of 20.

For  \res{the STFT-CNN}, the \gls*{tpr} of 91.5\% outperforms that of \res{the} QT\res{-CNN}, mainly \res{due to the} difficult instances of signals at very low \gls*{snr}. \res{It also gives a higher precision of 0.977, and a higher $F_1$-score of 0.945.} This \res{might reflect}  how using \res{a single} Q-range limits the frequency bandwidth of the input data, and how \res{a} fixed Q\res{-}range \res{might} not \res{be} optimal for all EOSs. On the one hand, the \gls*{stft} spectrogram data contain the full \res{frequency bandwidth of} CCSNe signal. On the other hand, the \res{QT-CNN} is penalized when using a single Q\res{-}range for all the EOSs, which limits the frequency bandwidth and \res{excludes} a significant \res{portion} of the \res{low and high} frequencies of the CCSNe signal. Given the current experimentation setup, STFT\res{-CNN} performs better than QT\res{-CNN} by detecting 8.6\% more event signals, using the default class threshold of $T=0.5$.  Most of the 8.6\% signals lie below  \res{\gls*{snr} $=0.5$} as shown in Figure~\ref{fig:test_set_accuracy}.

\begin{figure}
    \centering
    \includegraphics[width=1\linewidth]{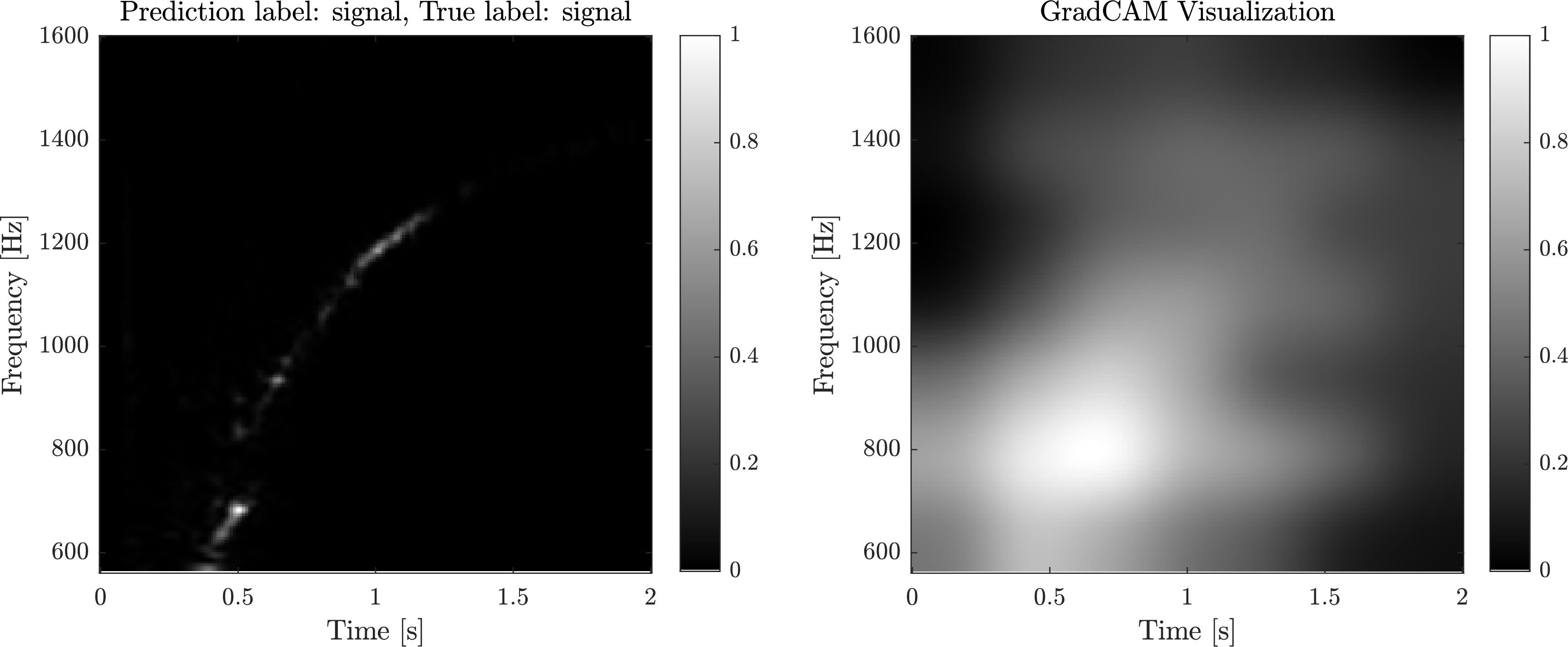}
    \caption{ \res{Left: QT spectrogram for a signal observation. Right: Grad-CAM visualization for the signal class prediction from the QT-CNN. The network focuses primarily on the region where the signal signature is most prominent.}}
    \label{fig:gradcam_qt}
\end{figure}

\begin{figure}
    \centering
    \includegraphics[width=1\linewidth]{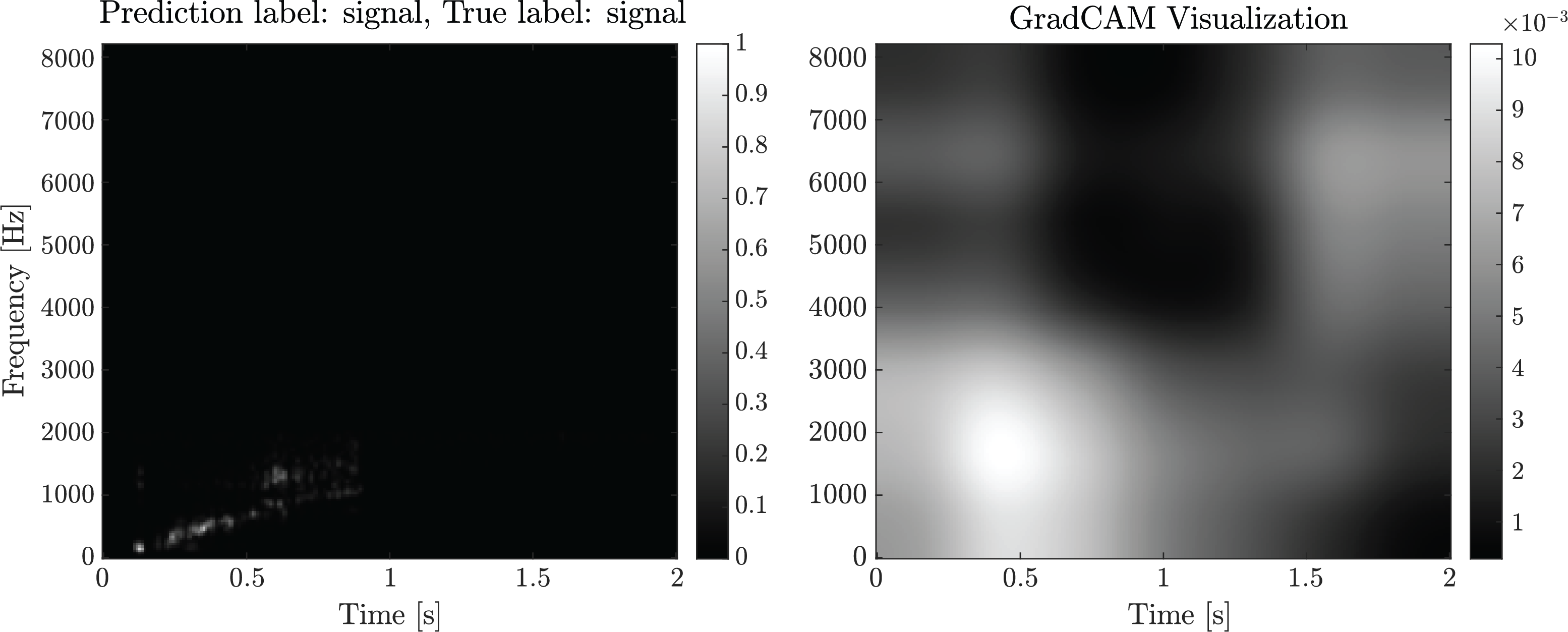}
    \caption{ \res{Left: STFT spectrogram for a signal observation. Right: Grad-CAM visualization for the signal class prediction from the STFT-CNN. The network concentrates most around the signal signature at 0.5 s and 1500 Hz.}}
    \label{fig:gradcam_stft}
\end{figure}

To better understand the network's decision process and ensure the classification made is justifiable, we examine the network with the Grad-CAM visualization \citep{Selvaraju2017}  \res{for one realization from the predictions of} both networks, shown in Figures~\ref{fig:gradcam_qt} and \ref{fig:gradcam_stft}. \res{This method gives an idea of which parts of the input image are most relevant to a prediction.} It reveals that the  \res{QT-CNN} focuses correctly at a blob centering around 0.6 second and 800 Hz, where the signal's time-frequency signature is most likely to appear.  Similar results are obtained for the  \res{STFT-CNN,} where the lower left region of the input contributes the most to the prediction of the signal class. \res{However, note that the area above 1600 Hz is relevant for the prediction, a region that was omitted in the QT-CNN and could explain the difference in their performance.}

\begin{table}
\centering
\caption{Confusion chart of the test set, positive class threshold $T=0.4$ for \gls*{qt}\res{-\gls*{cnn}} and $T=0.5$ for \gls*{stft}\res{-\gls*{cnn}}. The percentages show the proportion of positive/negative detections relative to the ground truth. The lowest \gls*{snr} in the test set is 0.2 from s20.0--SFHo at 10 kpc, as can  \res{be} seen in Figure~\ref{fig:snr_decay}.}
\label{tab:confusion-chart}
\begin{tblr}{
  width = \linewidth,
  colspec = {Q[87]Q[150]Q[165]Q[165]Q[165]Q[165]},
  row{4} = {c},
  row{5} = {c},
  cell{1}{2} = {c},
  cell{1}{3} = {c=4}{0.66\linewidth,c},
  cell{2}{3} = {c=2}{0.33\linewidth,c},
  cell{2}{5} = {c=2}{0.33\linewidth,c},
  cell{3}{2} = {c},
  cell{3}{3} = {c,b},
  cell{3}{4} = {c,b},
  cell{3}{5} = {c,b},
  cell{3}{6} = {c,b},
  cell{4}{1} = {r=2}{},
  hline{1,6} = {-}{0.08em},
  hline{3} = {3-6}{},
}
                                    &                 & Prediction    &                 &               &                 \\
                                    &                 &  \res{QT-CNN}            &                 &  \res{STFT-CNN}          &                 \\
                                    &                 & {\gls*{aligo}\\Noise} & {Event\\Signal} & {\gls*{aligo}\\Noise} & {Event\\Signal} \\
\begin{sideways}Truth\end{sideways} & {\gls*{aligo}\\Noise}   & 97.4 \%       & 2.6 \%          & 97.8 \%       & 2.2 \%          \\
                                    & {Event\\Signal} & 17.1 \%       & 82.9 \%         & 8.5 \%        & 91.5 \%         
\end{tblr}
\end{table}

\section{Conclusions}

Detecting \glspl*{gw} from \gls*{ccsne} could be the next breakthrough for ground-based detectors, underscoring the critical importance of advancing detection techniques.  In this paper, we have shown that a \gls*{cnn} can effectively detect these \gls*{gw} signals in the time-frequency domain.  We investigated two pre\res{-}processing methods for the time-frequency domain analysis \res{to train the \glspl*{cnn}}: \gls*{stft} and the \gls*{qt}.  Both approaches\res{, QT-CNN and STFT-CNN,} perform similarly in terms of noise and signal identification down to an \gls*{snr} of 0.5 \res{in our test set}. However,  \res{STFT-CNN} outperforms  \res{QT-CNN} in detecting signals below \res{an \gls*{snr} of} 0.5.  The \gls*{tpr}s are 0.83 and 0.92 for the  \res{QT-CNN} and  \res{STFT-CNN} for an \gls*{snr} as low as 0.2, respectively. \res{STFT-CNN also achieved a higher precision of 0.977, and a higher $F_1$-score of 0.945.}

Although  \res{STFT-CNN} yielded better results for faint signals \res{(those at extreme low \glspl*{snr})}, the experiment setup was designed for a singular \gls*{cnn} architecture to test both methods. This setup resulted in using one fixed Q\res{-}range to produce one \gls*{qt} spectrogram for each different data, which focuses on signal features in one portion of the whole frequency band. The choice of multiple Q\res{-}ranges to process each data will help cover the entire \gls*{ccsne} signal features and the whole frequency band. Therefore, a \gls*{cnn} architecture adapted for multiple \gls*{qt} spectrograms for each signal could potentially improve the \glspl*{cnn} performance in detecting  \glspl*{gw} from \gls*{ccsne}. 

\res{Our experiment was conducted using only simulated CCSNe GW and aLIGO noise signals. The performance of our \glspl*{cnn} in the context of real signals and real noise remains uncertain at this stage, partly because a CCSNe GW signal has not yet been observed. This represents a limitation of the study and an area for future work. However, it should be noted that using  aLIGO real noise data as pure noise for this analysis cannot be guaranteed due to various sources (e.g., environmental and instrumental factors), which could potentially introduce bias into the analysis, but may also reflect real-world complexities. Instead, a realistic aLIGO PSD \citep{aLIGOsens:2018} was used for this analysis, based on extensive observational data and modeling of noise sources in the aLIGO interferometers. Considering datasets from real aLIGO noise to test our proposed \glspl*{cnn} is part of our planned future work.}

\res{Our \glspl*{cnn}, like other similar algorithms, lack interpretability. To address this, we used Grad-CAM visualization to highlight the relevance of spectrogram regions in the prediction process, although the exact decision-making remains unclear. Nevertheless, \glspl*{cnn} are well-suited when prediction is the primary focus of the analysis. In particular, they are especially useful for CCSNe detection, as they do not require a theoretical background for training.}

\res{We have demonstrated that \glspl*{cnn} trained in the time-frequency domain are viable methods for CCSNe detection at low \glspl*{snr}. Our findings highlight the potential for exploring new techniques. Recent advances in neural-networks-based classification techniques (e.g., \cite{Song2024a, Song2024b, Abdullahi2024a, Abdullahi2024b}) present diverse opportunities for improving CCSNe detection.
 Similarly, the development of 3D simulations for CCSNe GW signals (e.g., \cite{choi:2024}) will provide more datasets and enable the training of \glspl*{cnn} in more realistic scenarios, improving detection accuracy. Additionally, a fast emulator for CCSNe signals \citep{Tarin:2024}, based on a deep convolutional generative adversarial network, has recently been proposed and may prove useful in developing advanced detection methods. The incorporation of   these new developments will play an integral role in future work. 
}

\section*{Acknowledgments} 

Authors' work was supported by the Collaborative International Interuniversity Research, Innovation, and Developement program-CIIRID: Continuity, Second Contest, from Pontificia Universidad Cat\'olica de Valpara\'iso, Chile, and Auckland University of Technology, New Zealand. PMR gratefully acknowledges support  by the Marsden Fund Council grant MFP-UOA2131 from New Zealand Government funding, managed by the Royal Society Te Apārangi.




\bibliographystyle{mnras}
\bibliography{reference} 

\begin{thebibliography}{}
\makeatletter
\relax
\def\mn@urlcharsother{\let\do\@makeother \do\$\do\&\do\#\do\^\do\_\do\%\do\~}
\def\mn@doi{\begingroup\mn@urlcharsother \@ifnextchar [ {\mn@doi@} {\mn@doi@[]}}
\def\mn@doi@[#1]#2{\def\@tempa{#1}\ifx\@tempa\@empty \href {http://dx.doi.org/#2} {doi:#2}\else \href {http://dx.doi.org/#2} {#1}\fi \endgroup}
\def\mn@eprint#1#2{\mn@eprint@#1:#2::\@nil}
\def\mn@eprint@arXiv#1{\href {http://arxiv.org/abs/#1} {{\tt arXiv:#1}}}
\def\mn@eprint@dblp#1{\href {http://dblp.uni-trier.de/rec/bibtex/#1.xml} {dblp:#1}}
\def\mn@eprint@#1:#2:#3:#4\@nil{\def\@tempa {#1}\def\@tempb {#2}\def\@tempc {#3}\ifx \@tempc \@empty \let \@tempc \@tempb \let \@tempb \@tempa \fi \ifx \@tempb \@empty \def\@tempb {arXiv}\fi \@ifundefined {mn@eprint@\@tempb}{\@tempb:\@tempc}{\expandafter \expandafter \csname mn@eprint@\@tempb\endcsname \expandafter{\@tempc}}}

\bibitem[\protect\citeauthoryear{Aasi et~al.,}{Aasi et~al.}{2015}]{aasi2015advanced}
Aasi J.,  et~al., 2015, \mn@doi [Classical and Quantum Gravity] {10.1088/0264-9381/32/7/074001}, 32, 074001

\bibitem[\protect\citeauthoryear{Abbott, Jawahar, Lockerbie  \& Tokmakov}{Abbott et~al.}{2016a}]{abbott2016ligo}
Abbott B.,  Jawahar S.,  Lockerbie N.,   Tokmakov K.,  2016a, \mn@doi [Physical Review D] {10.1103/PhysRevD.94.064035}, 94, 064035

\bibitem[\protect\citeauthoryear{Abbott et~al.,}{Abbott et~al.}{2016b}]{abbott2016gw150914}
Abbott B.~P.,  et~al., 2016b, \mn@doi [Physical Review Letters] {10.1103/PhysRevLett.116.131103}, 116, 131103

\bibitem[\protect\citeauthoryear{Abbott et~al.,}{Abbott et~al.}{2017a}]{Abbot2017}
Abbott B.~P.,  et~al., 2017a, \mn@doi [Physical Review Letters] {10.1103/PhysRevLett.119.161101}, 119, 161101

\bibitem[\protect\citeauthoryear{Abbott et~al.,}{Abbott et~al.}{2017b}]{abbott2017gravitational}
Abbott B.~P.,  et~al., 2017b, \mn@doi [The Astrophysical Journal Letters] {10.3847/2041-8213/aa920c}, 848, L13

\bibitem[\protect\citeauthoryear{Abdullahi, Chamnongthai, Gabralla  \& Chiroma}{Abdullahi et~al.}{2024a}]{Abdullahi2024a}
Abdullahi S.~B.,  Chamnongthai K.,  Gabralla L.~A.,   Chiroma H.,  2024a, \mn@doi [IEEE Sensors Journal] {10.1109/JSEN.2024.3407786}, 24, 37630

\bibitem[\protect\citeauthoryear{Abdullahi, Chamnongthai, Bolon-Canedo  \& Cancela}{Abdullahi et~al.}{2024b}]{Abdullahi2024b}
Abdullahi S.~B.,  Chamnongthai K.,  Bolon-Canedo V.,   Cancela B.,  2024b, \mn@doi [Expert Systems with Applications] {https://doi.org/10.1016/j.eswa.2024.123258}, 248, 123258

\bibitem[\protect\citeauthoryear{Agazie et~al.,}{Agazie et~al.}{2023}]{agazie2023nanograv}
Agazie G.,  et~al., 2023, \mn@doi [The Astrophysical Journal Letters] {10.3847/2041-8213/acdac6}, 951, L8

\bibitem[\protect\citeauthoryear{Antelis, Cavaglia, Hansen, Morales, Moreno, Mukherjee, Szczepa{\'n}czyk  \& Zanolin}{Antelis et~al.}{2022}]{antelis2022using}
Antelis J.~M.,  Cavaglia M.,  Hansen T.,  Morales M.~D.,  Moreno C.,  Mukherjee S.,  Szczepa{\'n}czyk M.~J.,   Zanolin M.,  2022, \mn@doi [Physical Review D] {10.1103/PhysRevD.105.084054}, 105, 084054

\bibitem[\protect\citeauthoryear{Astone, Cerd\'a-Dur\'an, Di~Palma, Drago, Muciaccia, Palomba  \& Ricci}{Astone et~al.}{2018}]{Astone2018}
Astone P.,  Cerd\'a-Dur\'an P.,  Di~Palma I.,  Drago M.,  Muciaccia F.,  Palomba C.,   Ricci F.,  2018, \mn@doi [Physical Review D] {10.1103/PhysRevD.98.122002}, 98, 122002

\bibitem[\protect\citeauthoryear{Bizouard, Maturana-Russel, Torres-Forn\'e, Obergaulinger, Cerd\'a-Dur\'an, Christensen, Font  \& Meyer}{Bizouard et~al.}{2021}]{Bizouard:2021}
Bizouard M.-A.,  Maturana-Russel P.,  Torres-Forn\'e A.,  Obergaulinger M.,  Cerd\'a-Dur\'an P.,  Christensen N.,  Font J.~A.,   Meyer R.,  2021, \mn@doi [Physical Review D] {10.1103/PhysRevD.103.063006}, 103, 063006

\bibitem[\protect\citeauthoryear{Brown}{Brown}{1991}]{brown1991calculation}
Brown J.~C.,  1991, \mn@doi [The Journal of the Acoustical Society of America] {10.1121/1.400476}, 89, 425

\bibitem[\protect\citeauthoryear{Chan, Heng  \& Messenger}{Chan et~al.}{2016}]{chan2020detection}
Chan M.~L.,  Heng I.~S.,   Messenger C.,  2016, \mn@doi [Physical Review D] {10.1103/PhysRevD.94.064035}, 94, 064035

\bibitem[\protect\citeauthoryear{Chatterji, Blackburn, Martin  \& Katsavounidis}{Chatterji et~al.}{2004}]{chatterji2004multiresolution}
Chatterji S.,  Blackburn L.,  Martin G.,   Katsavounidis E.,  2004, \mn@doi [Classical and Quantum Gravity] {https://doi.org/10.1088/0264-9381/21/20/024}, 21, S1809

\bibitem[\protect\citeauthoryear{Choi, Burrows  \& Vartanyan}{Choi et~al.}{2024}]{choi:2024}
Choi L.,  Burrows A.,   Vartanyan D.,  2024, Gravitational-Wave and Gravitational-Wave Memory Signatures of Core-Collapse Supernovae (\mn@eprint {arXiv} {2408.01525}), \url {https://arxiv.org/abs/2408.01525}

\bibitem[\protect\citeauthoryear{Danzmann \& R{\"u}diger}{Danzmann \& R{\"u}diger}{2003}]{danzmann2003lisa}
Danzmann K.,  R{\"u}diger A.,  2003, \mn@doi [Classical and Quantum Gravity] {10.1088/0264-9381/20/10/301}, 20, S1

\bibitem[\protect\citeauthoryear{Eccleston \& Edwards}{Eccleston \& Edwards}{2024}]{Tarin:2024}
Eccleston T.,  Edwards M.~C.,  2024, \mn@doi [Physical Review D] {10.1103/PhysRevD.110.104055}, 110, 104055

\bibitem[\protect\citeauthoryear{Einstein et~al.}{Einstein et~al.}{1916}]{einstein1916foundation}
Einstein A.,  et~al., 1916, \mn@doi [Annalen der Physik] {10.1002/andp.19163540702}, 49, 769

\bibitem[\protect\citeauthoryear{Fawcett}{Fawcett}{2006}]{FAWCETT2006861}
Fawcett T.,  2006, \mn@doi [Pattern Recognition Letters] {https://doi.org/10.1016/j.patrec.2005.10.010}, 27, 861

\bibitem[\protect\citeauthoryear{He, Zhang, Ren  \& Sun}{He et~al.}{2016}]{He2016}
He K.,  Zhang X.,  Ren S.,   Sun J.,  2016, in 2016 IEEE Conference on Computer Vision and Pattern Recognition (CVPR). pp 770--778, \mn@doi{10.1109/CVPR.2016.90}

\bibitem[\protect\citeauthoryear{Ioffe \& Szegedy}{Ioffe \& Szegedy}{2015}]{ioffe2015batch}
Ioffe S.,  Szegedy C.,  2015, Batch Normalization: Accelerating Deep Network Training by Reducing Internal Covariate Shift (\mn@eprint {arXiv} {1502.03167})

\bibitem[\protect\citeauthoryear{Kingma \& Ba}{Kingma \& Ba}{2017}]{kingma2017adam}
Kingma D.~P.,  Ba J.,  2017, Adam: A Method for Stochastic Optimization (\mn@eprint {arXiv} {1412.6980})

\bibitem[\protect\citeauthoryear{Klimenko, Yakushin, Mercer  \& Mitselmakher}{Klimenko et~al.}{2008}]{Klimenko_2008}
Klimenko S.,  Yakushin I.,  Mercer A.,   Mitselmakher G.,  2008, \mn@doi [Classical and Quantum Gravity] {10.1088/0264-9381/25/11/114029}, 25, 114029

\bibitem[\protect\citeauthoryear{Krishna, Devi, V  \& J}{Krishna et~al.}{2023}]{Music_CNN}
Krishna B.,  Devi G.~D.,  V S.,   J M.,  2023, in 2023 International Conference on System, Computation, Automation and Networking (ICSCAN). pp~1--6, \mn@doi{10.1109/ICSCAN58655.2023.10395616}

\bibitem[\protect\citeauthoryear{{Macleod}, {Areeda}, {Coughlin}, {Massinger}  \& {Urban}}{{Macleod} et~al.}{2021}]{gwpy}
{Macleod} D.~M.,  {Areeda} J.~S.,  {Coughlin} S.~B.,  {Massinger} T.~J.,   {Urban} A.~L.,  2021, \mn@doi [SoftwareX] {10.1016/j.softx.2021.100657}, 13, 100657

\bibitem[\protect\citeauthoryear{Matthew, Riccardo, Salvatore  \& Evan}{Matthew et~al.}{2018}]{aLIGOsens:2018}
Matthew E.,  Riccardo S.,  Salvatore V.,   Evan H.,  2018, \mn@doi [LIGO-T1500293-v13] {https://dcc.ligo.org/ligo-t1500293/public}

\bibitem[\protect\citeauthoryear{M{\"u}ller, Janka  \& Marek}{M{\"u}ller et~al.}{2013}]{muller2013new}
M{\"u}ller B.,  Janka H.-T.,   Marek A.,  2013, \mn@doi [The Astrophysical Journal] {https://doi.org/10.1088/0004-637X/766/1/43}, 766, 43

\bibitem[\protect\citeauthoryear{Nair \& Hinton}{Nair \& Hinton}{2010}]{Nair2010}
Nair V.,  Hinton G.~E.,  2010, in Proceedings of the 27th International Conference on International Conference on Machine Learning. ICML'10.
Omnipress, Madison, WI, USA, pp 807--814

\bibitem[\protect\citeauthoryear{Nunes, Escrig, Freitas, Font, Fernandes, Onofre  \& Torres-Forn{\'e}}{Nunes et~al.}{2024}]{nunes2024deep}
Nunes S.,  Escrig G.,  Freitas O.~G.,  Font J.~A.,  Fernandes T.,  Onofre A.,   Torres-Forn{\'e} A.,  2024, Deep-Learning Classification and Parameter Inference of Rotational Core-Collapse Supernovae (\mn@eprint {arXiv} {2403.04938})

\bibitem[\protect\citeauthoryear{Sassi, Haleem  \& Pecchia}{Sassi et~al.}{2024}]{Sassi2024}
Sassi M.,  Haleem M.~S.,   Pecchia L.,  2024, Spectrogram-Based Approach with Convolutional Neural Network for Human Activity Classification.
Springer Nature Switzerland, pp 387--401, \mn@doi{10.1007/978-3-031-49068-2_40}

\bibitem[\protect\citeauthoryear{Selvaraju, Cogswell, Das, Vedantam, Parikh  \& Batra}{Selvaraju et~al.}{2017}]{Selvaraju2017}
Selvaraju R.~R.,  Cogswell M.,  Das A.,  Vedantam R.,  Parikh D.,   Batra D.,  2017, in 2017 IEEE International Conference on Computer Vision (ICCV). pp 618--626, \mn@doi{10.1109/ICCV.2017.74}

\bibitem[\protect\citeauthoryear{Song, Wu, Song, Stojanovic  \& Tejado}{Song et~al.}{2024a}]{Song2024a}
Song X.,  Wu C.,  Song S.,  Stojanovic V.,   Tejado I.,  2024a, \mn@doi [Engineering Applications of Artificial Intelligence] {https://doi.org/10.1016/j.engappai.2023.107832}, 131, 107832

\bibitem[\protect\citeauthoryear{Song, Peng, Song  \& Stojanovic}{Song et~al.}{2024b}]{Song2024b}
Song X.,  Peng Z.,  Song S.,   Stojanovic V.,  2024b, \mn@doi [Communications in Nonlinear Science and Numerical Simulation] {https://doi.org/10.1016/j.cnsns.2024.107945}, 132, 107945

\bibitem[\protect\citeauthoryear{Stockwell, Mansinha  \& Lowe}{Stockwell et~al.}{1996}]{492555}
Stockwell R.,  Mansinha L.,   Lowe R.,  1996, \mn@doi [IEEE Transactions on Signal Processing] {10.1109/78.492555}, 44, 998

\bibitem[\protect\citeauthoryear{Vartanyan, Burrows, Wang, Coleman  \& White}{Vartanyan et~al.}{2023}]{vartanyane3d}
Vartanyan D.,  Burrows A.,  Wang T.,  Coleman M. S.~B.,   White C.~J.,  2023, \mn@doi [Phys. Rev. D] {10.1103/PhysRevD.107.103015}, 107, 103015

\bibitem[\protect\citeauthoryear{Vink}{Vink}{2020}]{supernovaebook}
Vink J.,  11 November 2020, Physics and Evolution of Supernova Remnants.
Springer Cham, \mn@doi{10.1007/978-3-030-55231-2}

\bibitem[\protect\citeauthoryear{Wang, Zhuang, Tao, Paszke  \& Stojanovic}{Wang et~al.}{2023}]{Wang2023}
Wang R.,  Zhuang Z.,  Tao H.,  Paszke W.,   Stojanovic V.,  2023, \mn@doi [ISA Transactions] {https://doi.org/10.1016/j.isatra.2023.07.043}, 142, 123

\bibitem[\protect\citeauthoryear{Wolfe, Fr{\"o}hlich, Miller, Torres-Forn{\'e}  \& Cerd{\'a}-Dur{\'a}n}{Wolfe et~al.}{2023}]{wolfe2023gravitational}
Wolfe N.~E.,  Fr{\"o}hlich C.,  Miller J.~M.,  Torres-Forn{\'e} A.,   Cerd{\'a}-Dur{\'a}n P.,  2023, \mn@doi [The Astrophysical Journal] {10.3847/1538-4357/ace693}, 954, 161

\makeatother
\end{thebibliography}








\bsp	
\label{lastpage}


\end{document}